\documentclass[11pt,english,floatfix,nofootinbib,superscriptaddress,aps,prd,preprint]{revtex4}
\usepackage[utf8]{inputenc}
\usepackage{float}
\usepackage{array}
\usepackage{lipsum}
\usepackage{graphicx}
\usepackage{amsmath,amsthm,amsfonts,amssymb}
\usepackage{graphicx}
\usepackage[english]{babel}
\usepackage{xcolor}
\usepackage{tensor}
\usepackage{esint}
\usepackage[dvips]{epsfig}
\usepackage[dvips]{graphicx}
\usepackage{float}
\usepackage{units}
\usepackage{textcomp}
\usepackage{mathrsfs}
\usepackage{amsmath}
\usepackage[makeroom]{cancel}
\usepackage{amssymb}
\usepackage{amsbsy}
\usepackage{amsfonts}
\usepackage{amssymb,mathrsfs,xcolor}
\usepackage{esint}
\usepackage{array}
\usepackage{graphicx}
\usepackage{mathtools,slashed}
\usepackage{slashed}
\usepackage{hyperref}
\hypersetup{
    colorlinks,
    citecolor=blue,
    filecolor=green,
    linkcolor=purple,
    urlcolor=red,
}

\newcommand{\ie}{\begin{equation}}
\newcommand{\fe}{\end{equation}}
\newcommand{\se}{\begin{eqnarray}}
\newcommand{\ff}{\end{eqnarray}}

\newcommand{\dd}{\mathrm{d}}

\usepackage{hyperref}
\hypersetup{colorlinks,breaklinks,
			citecolor=[rgb]{0,0.0,1.0},
            urlcolor=[rgb]{0.0,0.0,1.0},
            linkcolor=[rgb]{0,0.5,0.9}}

\begin{document}

\title{Nonrotating and rotating black holes with secondary disformal hair in a ghost-free metric-affine parity-violating scalar-torsion theory}

\author{E. Brito}
\email{eliasbaj@ufob.edu.br}
\affiliation{Centro de Ci\^encias Exatas e das Tecnologias, Universidade Federal do Oeste da Bahia - Rua Bertioga 892, 47810-059, Barreiras,
BA, Brazil}
\affiliation{Departamento de Física, Universidade Federal da Paraíba, Caixa Postal 5008, 58051-970, João Pessoa, Paraíba,  Brazil} 

\author{G. Macedo}
\email{gabriel.santana@academico.ufpb.br}
\affiliation{Departamento de Física, Universidade Federal da Paraíba, Caixa Postal 5008, 58051-970, João Pessoa, Paraíba,  Brazil}

\author{J. R. Nascimento}
\email{jroberto@fisica.ufpb.br}
\affiliation{Departamento de Física, Universidade Federal da Paraíba, Caixa Postal 5008, 58051-970, João Pessoa, Paraíba,  Brazil} 

\author{A. Yu. Petrov}
\email{petrov@fisica.ufpb.br}
\affiliation{Departamento de Física, Universidade Federal da Paraíba, Caixa Postal 5008, 58051-970, João Pessoa, Paraíba,  Brazil} 

\author{P. J. Porfírio}
\email{pporfirio@fisica.ufpb.br}
\affiliation{Departamento de Física, Universidade Federal da Paraíba, Caixa Postal 5008, 58051-970, João Pessoa, Paraíba,  Brazil}


\date{\today}

\begin{abstract}
In this paper, we formulate a ghost-free metric-affine scalar-torsion model in which a
pseudoscalar field is coupled to the Nieh--Yan density and derivatively coupled to the
Einstein-Palatini tensor. The theory is projectively invariant and, in the shift-symmetric
sector, depends on the scalar field only through its gradient. After fixing the projective gauge,
the independent connection is solved exactly: its symmetric part is the Levi-Civita
connection associated with the metric $h_{\mu\nu}$, disformally related to the metric $g_{\mu\nu}$, while its antisymmetric part is a purely
axial torsion sourced by $v_\mu=\partial_\mu\vartheta$.  We identify a degenerate ghost-free branch,
$\alpha=6\beta^2\kappa^2$, for which Ricci-flat geometries in the $h$-metric frame solve
the metric equations and the Noether current vanishes identically. For example, we consider
Schwarzschild and Kerr
geometries in the $h$-metric frame, which generate disformal deformations on these geometries from the perspective of the $g$-metric frame. For the Schwarzschild black hole in the $h$-metric frame, we construct static and
Babichev--Charmousis-like time-dependent scalar profiles, and discuss horizon regularity in
Eddington--Finkelstein coordinates. For the Kerr black hole in the $h$-metric frame, we show that regularity on the rotation axis rules out a nontrivial globally real static fixed-norm scalar profile. For the stationary time-dependent branch, axis regularity and asymptotic reality fix $Y_0=-q^2$ and eliminate the angular dependence, yielding a nontrivial future-regular profile in ingoing Kerr coordinates. The Kerr solution with the nontrivial scalar field 
 is stealth only in the $h$-metric frame. In the $g$-metric frame, it produces
 disformal corrections stemming from the nontrivial scalar field profile and it represents secondary disformal scalar hair rather
than primary Noether-charge hair.
\end{abstract}

\maketitle

\section{Introduction}
\label{sec:intro}

Undoubtedly, General Relativity (GR) has been remarkably successful over the last century. Its theoretical predictions have been confirmed by numerous experiments to date -- in particular, a notable recent example is the observation of black-hole shadows by
 the Event Horizon Telescope Collaboration \cite{EHT:2019dse, EventHorizonTelescope:2022wkp} (for a review on black hole shadows, see \cite{Vagnozzi:2022moj}) and detection of gravitational waves \cite{LIGOScientific:2016lio}. Despite the astounding success of GR, some issues remain difficult to consistently explain within this framework. For example, astrophysical observational data \cite{Planck:2018vyg} suggest that our Universe is almost entirely composed of exotic matter/energy sources (dark matter and energy) which are difficult to reconcile with GR. Given this difficulty (as well as other ones, see e.g. \cite{Clifton:2011jh}), a variety of attempts have been proposed in the literature in order to resolve these problems consistently. Many of them are based on modifications of GR, or alternative theories of gravity, obtained by including new dynamical fields (non-)minimally coupled to gravity, such as, for example, Brans-Dicke and Horndeski theories \cite{Joyce:2014kja,Horndeski:1974wa,Zumalacarregui:2013pma}, or by including higher-order curvature terms
 in the gravitational sector see, e.g. \cite{Astashenok:2017dpo}.  On the other hand, the existence of a large number of alternative theories of gravity (see, for a review, e.g. \cite{Odintsov1,Odintsov2,Harko:2018ayt,Petrov:2020wgy}) could be, in some sense, an additional difficulty in identifying the most appropriate one.  

Scalar-tensor theories are particularly interesting models of modified theories of gravity. They provide an important arena for testing the robustness of black-hole no-hair arguments. A standard current-based no-hair theorem for shift-symmetric Galileons \cite{Nicolis:2008in} was formulated by Hui and Nicolis~\cite{Hui:2012qt}. Babichev and Charmousis later showed that such arguments can be evaded by allowing scalar profiles with linear time dependence while keeping the metric static, yielding regular black holes with nontrivial scalar configurations~\cite{Babichev:2013cya}. Black hole solutions have been developed within scalar-tensor theories \cite{Rinaldi:2012vy} and, in particular, in Horndeski theories \cite{Charmousis:2011bf}. In particular, the regularity of the scalar must be checked in coordinates that are regular at the future horizon, such as advanced Eddington--Finkelstein coordinates. Such configurations are usually referred to as stealth black holes (Ricci-flat metrics sourced by a non-trivial scalar field (hair) \cite{Babichev:2013cya, Bakopoulos:2023fmv}).
Black-hole hair is usually classified as primary or secondary: primary hair corresponds to an independent conserved charge associated with the nontrivial scalar field, whereas secondary hair is fixed by the usual global charges, namely mass ($M$), charge ($Q$), and angular momentum ($J$).

Degenerate Higher-Order Scalar-Tensor Theories (DHOST) provide a broad framework for constructing higher-derivative scalar-tensor models that avoid the propagation of undesirable ghost degrees of freedom. They generalize Horndeski theories by allowing Lagrangians containing higher-order derivative interactions between the metric and the scalar field, while maintaining a degenerate kinetic structure. This degeneracy removes the would-be Ostrogradsky mode and ensures that the theory propagates the correct number of physical degrees of freedom \cite{Langlois:2015cwa, Langlois:2018dxi}. In this sense, DHOST offer a systematic way of extending scalar-tensor gravities beyond the Horndeski class while preserving ghost freedom. Furthermore, DHOST coupled with topological invariants \cite{Chern,Zanelli,Naka} have attracted considerable attention, mainly in the context of Lorentz/CPT symmetry breaking (LSB), for example, the Chern-Simons modified gravity (CSMG) \cite{Jackiw:2003pm,Alexander:2009tp,Boudet:2022wmb} characterized by a (pseudo-)scalar field coupled with the gravitational Chern-Pontryagin topological invariant \cite{Witten,Deser}. The CSMG itself is an example of a parity-violating (CPT-violating) model (and, for a certain form of the CS coefficient, a Lorentz-violating one)  \cite{Jackiw:2003pm,Mirza, Bartolo}. Parity-violating gravity effects have been extensively studied in the scenario of gravitational waves \cite{Cook,Yunes1,Alexander2,Conroy,Nishi,Zhao1,Qiao,Qiao2,Wang}. On top of that, early Universe cosmological parity-violating effects and their imprints have been explored in the literature \cite{Lue,Alexander1,Saito,Sorbo,Soda,Bartolo1,Phil,Phil1}. In the context of astrophysics, black hole perturbations in parity-violating theories have also been investigated \cite{Moto,Taha,Taha1,Lin,Yu:2025fcv}, as well as G\"{o}del-type metrics \cite{Nascimento}.

 The aforementioned scalar-tensor theories have mostly been explored within the metric approach, in which the only dynamical quantities are the metric and the scalar field. A possible natural way to generalize them is to consider such theories within the metric-affine (Palatini) approach, a formalism in which the metric and the connection are treated as independent variables a priori \cite{Olmo:2011uz,Olmo-Garcia,Eisenhart}.  Torsion and nonmetricity are therefore present before the connection equation is solved. Parity-violating topological densities, such as the Nieh--Yan invariant \cite{Nieh,Bomba}, are naturally constructed from the torsion tensor.

The purpose of this paper is to construct a ghost-free metric-affine parity-violating model coupling a topological invariant based on the torsion tensor with a (pseudo-)scalar field. The Nieh--Yan invariant is a natural candidate because it is intrinsically tied to torsion and becomes nontrivial only in torsional geometries. Furthermore, by requiring projective invariance, the connection equation can be solved exactly. The symmetric part becomes the Levi-Civita connection of the metric $h_{\mu\nu}$, which is disformally related to $g_{\mu\nu}$, while the torsion is purely axial and sourced by $v_\mu=\partial_\mu\vartheta$. 

The central result is that, for the following choice of parameters,
\begin{equation}
    \alpha=6\beta^2\kappa^2,
    \label{eq:intro-branch}
\end{equation}
Ricci-flat geometries solve the metric equation in the $h$-metric frame. Thus, Schwarzschild and Kerr metrics in the $h$-frame generate exact black-hole solutions. The scalar is stealth with respect to $h_{\mu\nu}$, because the geometry remains a vacuum solution in the $h$-frame despite the nontrivial scalar. However, the physical metric $g_{\mu\nu}$ is generally deformed by the scalar gradient through the disformal map. The resulting scalar should therefore be interpreted as secondary disformal scalar hair, not as primary Noether-charge hair.

The paper is organized as follows. In Section \ref{sec2}, we present the ghost-free metric-affine parity-violating model. We obtain the field equations, formally solve the connection equation, and demonstrate, as a result of the projective invariance, that the scalar field equation reduces to a conserved current equation. In Section \ref{sec3}, we obtain (non-)rotating black hole solutions for nontrivial scalar field profiles. In Section \ref{conc}, we present our conclusions.

We use the following conventions throughout the paper: the mostly-plus signature $(-,+,+,+)$, $\kappa^2=8\pi G$, the Riemann tensor is defined by $R^{\mu}_{\phantom{a}\nu\alpha\beta}=\partial_{\alpha}\Gamma^{\mu}_{\beta\nu}+\Gamma^{\mu}_{\alpha\lambda}\Gamma^{\lambda}_{\beta\nu}-(\alpha\longleftrightarrow\beta)$, and 
$A_{(\mu\nu)}=\frac12(A_{\mu\nu}+A_{\nu\mu})$ and
   $ A_{[\mu\nu]}=\frac12(A_{\mu\nu}-A_{\nu\mu})$ for symmetrization and antisymmetrization tensors, respectively.

\section{Ghost-free metric-affine parity-violating modified gravity}
\label{sec2}

Our starting point will be to define the proposed modified theory of gravity. The model basically consists of adding two nonminimal couplings -- one of which involves a (pseudo)scalar field ($\vartheta$) coupled to the Nieh-Yan density and the other one involving derivatives of the scalar field coupled to the Ricci scalar -- to the Einstein-Palatini action. Now, let us write the action explicitly 
\begin{equation}
S=S_{EP}+S_{NM}+S_{\vartheta}+S_{m},
\label{total}
\end{equation}    
where the Einstein-Palatini action is defined by
\begin{equation}
S_{EP}=\frac{1}{2\kappa^2}\int d^{4}x\sqrt{-g}\, R(\Gamma),
\end{equation}
where $R(\Gamma)=g^{\mu\nu}R_{\mu\nu}(\Gamma)$ is the Ricci scalar and $R_{\mu\nu}(\Gamma)$ is the Ricci tensor, both defined in the Palatini framework, and the non-minimal couplings are given by
\begin{equation}
S_{NM}=\beta\int d^{4}x\sqrt{-g}\,\vartheta NY + \frac{\gamma}{4\kappa^2}\int d^{4}x\sqrt{-g}\,\left(X R(\Gamma)+2g^{\mu\alpha}g^{\nu\beta}(\partial_{\alpha}\vartheta)(\partial_{\beta}\vartheta)G_{\mu\nu}(\Gamma)\right),  
\label{NY}
\end{equation}
where $X\equiv g^{\mu\nu}(\partial_{\mu}\vartheta)(\partial_{\nu}\vartheta)$. The first term in the action above stands for the non-minimal coupling between the topological Nieh-Yan term and the scalar field, while the second one is the coupling between the Einstein-Palatini tensor, $G_{\mu\nu}(\Gamma)=R_{\mu\nu}(\Gamma)-\frac{1}{2}g_{\mu\nu}R(\Gamma)$, with first-order derivatives of the scalar field. In addition, $\beta$ and $\gamma$ are coupling constants controlling the non-minimal interactions. The Nieh-Yan term is defined by
\begin{equation}
NY\equiv\epsilon^{\mu\nu\alpha\beta}\left(S^{\rho}_{\,\,\mu\nu}(S_{\rho\alpha\beta}+Q_{\alpha\beta\rho})-\dfrac{1}{2}R_{\mu\nu\alpha\beta}\right),
\end{equation} 
where $\epsilon^{\mu\nu\alpha\beta}=\dfrac{\varepsilon^{\mu\nu\alpha\beta}}{\sqrt{-g}}$ is the Levi-Civita tensor and $\epsilon^{\mu\nu\alpha\beta}R_{\mu\nu\alpha\beta}$ is the Holst term \cite{Holst:1995pc}. The minimal scalar sector is defined by  
\begin{equation}
S_{\vartheta}=\alpha\int d^{4}x\sqrt{-g}\bigg[\frac{1}{2}(\partial_{\mu}\vartheta)(\partial^{\mu}\vartheta)-V(\vartheta)\bigg],
\end{equation}
 we have introduced the real constant parameter $\alpha$ for the sake of future convenience, the special case $\alpha=1$ captures the standard scalar action, while the case $\alpha<0$ describes scalar fields with ``wrong'' sign kinetic term (ghost field). Finally, the contribution coming from matter sources is defined by
\begin{equation}
S_{m}=\int d^{4}x\sqrt{-g}\,\mathcal{L}_{m}(g_{\mu\nu}, \Gamma^{\mu}_{\alpha\beta}, \Psi),
\end{equation}
where $\Psi$ stands for matter sources different from $\vartheta$.
Some remarks should be made on Eq.(\ref{NY}). First, the Nieh-Yan term is a topological invariant, which means that it can be rewritten as the divergence of a vector (see, for example, \cite{Nieh})
\begin{equation}
NY=-\nabla^{(g)}_{\mu}\bar{S}^{\mu},
\end{equation} 
where $\bar{S}^{\mu}=\epsilon^{\mu\nu\alpha\beta}S_{\nu\alpha\beta}$ is the axial piece of the torsion tensor. Taking this into account, $S_{NY}$ can be expressed as
\begin{equation}
S_{NM}=\beta\int d^{4}x \sqrt{-g}\, v_{\mu}\bar{S}^{\mu} + \frac{\gamma}{4\kappa^2}\int d^{4}x\sqrt{-g}\,\left(X R(\Gamma)+2g^{\mu\alpha}g^{\nu\beta}(\partial_{\alpha}\vartheta)(\partial_{\beta}\vartheta)G_{\mu\nu}(\Gamma)\right),
\end{equation} 
where we have integrated Eq.(\ref{NY}) by parts and disregarded boundary terms and  defined $v_{\mu}=\nabla^{(g)}_{\mu}\vartheta=\partial_{\mu}\vartheta$. The first term in the action $S_{NM}$, written in the previous shape, represents by itself a contact interaction between two axial vectors: $v_{\mu}$ and $\bar{S}^{\mu}$. 

In analogy with Chern-Simons modified gravity \cite{Jackiw:2003pm}, one can lay out two different formulations:
\begin{itemize}
	\item Non-dynamical formulation ($\alpha=0$ and $\gamma=0$). It occurs when $\vartheta$ is taken to be a background field which entails that $v_{\mu}$ can be interpreted as a local Lorentz-violating background
in such a way that this coupling matches one of the possible terms proposed in the
Standard-Model Extension (SME) \cite{Kostelecky:2003fs}. As a result, we shall see that the theory within this formulation must fulfill a constraint in just the same way as the Pontryagin one in Chern-Simons modified gravity \cite{Jackiw:2003pm}. Moreover, requiring $\vartheta=\mbox{const.}$ which is the same as taking $v_{\mu}=0$  leads to the vanishing of $S_{NY}$ and $S_{\vartheta}$ and then the model reduces to the Einstein-Palatini with matter sources. 
\end{itemize}

\begin{itemize}
	\item Dynamical formulation ($\alpha\neq 0$ and $\gamma\neq 0$). In this case, $\vartheta$ is treated as a fundamental field of the model, and is therefore described by its own equation of motion rather than a constraint. 
\end{itemize}
 Another striking feature is related to the parity symmetry; notice that the Nieh-Yan term is odd under parity transformation since it is defined in terms of the Levi-Civita tensor; as a result, $\vartheta$ must also be odd in order to maintain the invariance of the full action under parity transformations. Conversely, if $\vartheta$ is even, the model breaks the parity invariance explicitly. 

The full action (\ref{total}), following the definitions outlined above, takes the explicit form
\begin{eqnarray}
S&=&\int d^{4}x \sqrt{-g} \left\{\frac{1}{2\kappa^2}\left[R(\Gamma)\left(1+\frac{\gamma}{2}X\right)+\gamma v^{\mu}v^{\nu}G_{\mu\nu}(\Gamma)\right]+\beta v_{\mu}\bar{S}^{\mu}+\alpha\bigg[\frac{1}{2}X-V(\vartheta)\bigg]+\right.\nonumber\\ &+&\left. \mathcal{L}_{m}\right\}.
\label{full}
\end{eqnarray} 

Let us now turn our attention to other symmetries of this action: 

\begin{enumerate}
	\item For a trivial potential ($V(\vartheta)=0$),
      it is straightforward to check that Eq.(\ref{full}) is invariant (up to boundary terms\footnote{We shall focus on the equations of motion of the action \eqref{full}, thus boundary terms can be neglected in our analysis.}) under global shift transformation of the scalar, \textit{i.e.}, $\vartheta\rightarrow\vartheta+a$, with $a$ being an arbitrary constant. Such a global symmetry, as usual, leads to a conserved Noether current $J^{\mu}$ that we shall define later. 
	\item For a nontrivial potential ($V(\vartheta)\neq 0$), the shift invariance is broken down due to the presence of non-derivative terms (self-interaction terms) in the scalar sector of the model.
\end{enumerate}

In addition, the action (\ref{full}) -- regardless of the potential shape -- is invariant under projective transformation,
\begin{equation}
\Gamma^{\alpha}_{\mu\nu}\longrightarrow \Gamma^{\alpha}_{\mu\nu}+\delta^{\alpha}_{\mu}A_{\nu},
\label{proj}
\end{equation}
where $A_{\mu}$ stands for a non-physical arbitrary vector field. Under this projective transformation, the Riemann tensor transforms as
\begin{equation}
R^{\mu}_{\phantom{a}\nu\alpha\beta}\longrightarrow R^{\mu}_{\phantom{a}\nu\alpha\beta}-2\delta^{\mu}_{\nu}\partial_{[\alpha}A_{\beta]}.
\end{equation}
It was argued in \cite{BeltranJimenez:2019acz} that terms breaking the projective symmetry would be connected with the possibility of the emergence of instabilities in higher-order metric-affine curvature theories. Nevertheless, our modified model contains only the symmetric piece of the Ricci tensor and the axial piece of the torsion tensor, which are invariant under projective transformations; as a result, the action is also invariant under these transformations. Such an invariance is typical in Ricci-based theories of gravity \cite{Afonso:2018hyj}, which avoid the emergence of ghost instabilities in the gravity sector. 

From now on, we shall concentrate our attention on the shift/projective-symmetric model, which corresponds to getting rid of the potential term in Eq.(\ref{full}), so we put $V(\vartheta)=0$ henceforth.

\subsection{Field equations}

We start this section by varying the action (\ref{full}) with respect to the metric, connection, and the scalar field, respectively, to find the field equations
\begin{align}
\nonumber & R_{(\mu\nu)}(\Gamma)-\frac{1}{2}g_{\mu\nu}\left(R(\Gamma)+\gamma v^{\alpha}v^{\beta}R_{\alpha\beta}(\Gamma)\right)+2\gamma v^{\beta}v_{(\mu}R_{\nu)\beta}(\Gamma)-2\beta\kappa^{2}v_{\lambda}\epsilon^{\lambda}_{\,\,\,\beta\gamma(\mu}S_{\nu)}^{\,\,\,\beta\gamma}\\
& =\kappa^{2}T_{\mu\nu};\label{me}\\
 \nonumber & \frac{\delta^{\nu}_{\alpha}}{\sqrt{-h}}\nabla^{(\Gamma)}_{\rho}\left[\sqrt{-h}h^{\mu\rho}\right]-\frac{1}{\sqrt{-h}}\nabla^{(\Gamma)}_{\alpha}\left[\sqrt{-h}h^{\mu\nu}\right]+2 S^{\nu}_{\alpha\rho}h^{\mu\rho}+2 S^{\sigma}_{\sigma\alpha}h^{\nu\mu}-\\
&-2\delta^{\nu}_{\alpha}S^{\sigma}_{\sigma\rho}h^{\rho\mu}+2\beta\kappa^{2} v_{\lambda}\frac{\varepsilon^{\lambda\vartheta\nu\mu}}{\sqrt{-h}}g_{\alpha\vartheta}=\frac{\kappa^{2}}{\sqrt{1+\gamma X}}\Delta_{\alpha}^{\nu\mu};\label{con}\\
& \alpha\square^{(g)}\vartheta+\frac{\gamma}{\kappa^2}\nabla^{(g)}_{\mu}\left[v_{\sigma}R^{(\mu\sigma)}(\Gamma)\right]=-\beta\nabla_{\mu}^{(g)}\bar{S}^{\mu}\label{sc},
\end{align}
  where we have defined the following quantities: the $h_{\mu\nu}$ metric, sometimes called the auxiliary metric (it is introduced as in \cite{Delhom:2019wcm}):
  \begin{equation}
   \sqrt{-h}h^{\mu\nu}=\sqrt{-g}(g^{\mu\nu}+\gamma v^{\mu}v^{\nu}),
   \label{hg}
   \end{equation}
  the torsion, which is, as usual, the antisymmetric part of the connection, $S^{\alpha}_{\mu\nu}=\Gamma^{\alpha}_{[\mu\nu]}$, and the stress-energy tensor $T_{\mu\nu}$ which might be split into two parts corresponding to the usual matter and $\vartheta$ field respectively: $T_{\mu\nu}=T^{(m)}_{\mu\nu}+T^{(\vartheta)}_{\mu\nu}$, with   
\begin{align}
T^{(m)}_{\mu\nu}&=-\frac{2}{\sqrt{-g}}\frac{\delta S_{m}}{\delta g^{\mu\nu}};\\	
\nonumber T^{(\vartheta)}_{\mu\nu}&=-\frac{2}{\sqrt{-g}}\frac{\delta S_{\vartheta}}{\delta g^{\mu\nu}}=-\alpha\left[(\partial_{\mu}\vartheta)(\partial_{\nu}\vartheta)-\frac{1}{2}g_{\mu\nu}(\partial_{\alpha}\vartheta)(\partial^{\alpha}\vartheta)\right],
\label{Tth}
\end{align}
and finally, the hypermomentum is given by
	\begin{equation}
	\Delta_{\alpha}^{\mu\nu}=-\frac{2}{\sqrt{-g}}\frac{\delta S_{m}}{\delta \Gamma^{\alpha}_{\mu\nu}}.
	\end{equation}
Hereafter we concentrate our analysis on pure scalar-tensor interactions, which amounts to taking $S_{m}=0$ or, equivalently, $\Delta_{\alpha}^{\mu\nu}=T_{\mu\nu}^{(m)}=0$.

Getting back to the metric equation (\ref{me}), it is convenient for latter purposes to rewrite it in the trace-reversed form. To do so, let us contract Eq.(\ref{me})  with the metric $g^{\mu\nu}$,
\begin{equation}
R(\Gamma)=-\kappa^2 (T+2\beta v_{\lambda}\bar{S}^{\lambda}),
\label{eqrw}
\end{equation}
which relates the Ricci scalar with the trace of the stress-energy tensor ($T=g^{\mu\nu}T_{\mu\nu}$) and the projection of the vector $v_{\mu}$ along the axial vector $\bar{S}_{\mu}$. In particular, as both vectors are mutually orthogonal, Eq.(\ref{eqrw}) reduces to a form similar to GR. Now, contracting Eq.(\ref{me}) with $v^{\mu}$ and $v^{\mu}v^{\nu}$, we are able to obtain after some algebraic manipulations that
\begin{eqnarray}
\nonumber v^{\mu}R_{(\mu\nu)}(\Gamma)&=&\frac{\kappa^2}{(1+\gamma X)}\bigg(v^{\mu}T_{\mu\nu}-\frac{\gamma v_{\nu}v^{\mu}v^{\lambda}T_{\mu\lambda}+(1+\gamma X)v_{\nu}T+2(1+\gamma X)\beta v_{\nu}v_{\lambda}\bar{S}^{\lambda}}{(2+3\gamma X)}+\\ \label{eqrw1}&+&\beta v_{\lambda}v_{\mu}\epsilon^{\lambda}_{\,\,\beta\gamma\nu}S^{\mu\beta\gamma}\bigg);\\
\label{eqrw2}v^{\mu}v^{\nu}R_{\mu\nu}(\Gamma)&=&\frac{\kappa^2}{(2+3\gamma X)}\left(2v^{\mu}v^{\nu}T_{\mu\nu}-XT-2\beta X v_{\lambda}\bar{S}^{\lambda}\right).
\end{eqnarray}
  Inserting Eqs.(\ref{eqrw}, \ref{eqrw1}) and (\ref{eqrw2}) into Eq.(\ref{me}), one gets
  \begin{eqnarray}
  \nonumber R_{(\mu\nu)}(\Gamma)&=&\kappa^2 T_{\mu\nu} +\frac{\kappa^2}{2}g_{\mu\nu}\left[\frac{\gamma}{(2+3\gamma X)}\left(2v^{\mu}v^{\nu}T_{\mu\nu}-XT-2\beta X v_{\lambda}\bar{S}^{\lambda}\right)-T\right]-\\
  \nonumber &-&\frac{2\gamma\kappa^2}{(1+\gamma X)}\bigg[v^{\beta}v_{(\mu}T_{\nu)\beta}+\beta v_{\lambda}v_{\vartheta}S^{\vartheta\beta\gamma}v_{(\mu}\epsilon^{\lambda}_{\,\,\nu)\beta\gamma}-\frac{2v_{\mu}v_{\nu}}{(2+3\gamma X)}
\bigg(\gamma v^{\vartheta}v^{\lambda}T_{\vartheta\lambda}+\\
&+&(1+\gamma X)T+2(1+\gamma X)\beta v_{\lambda}\bar{S}^{\lambda}\bigg)  \bigg]+2\beta \kappa^2 v_{\lambda}\epsilon^{\lambda}_{\,\,\beta\gamma(\mu}S_{\nu)}^{\,\,\beta\gamma}.
\label{mest}
  \end{eqnarray}
 Note that this equation relates the connection with the vector $v_{\mu}$ in an unusual way. Nevertheless, it is worth emphasizing the presence of nonminimal terms, such as $v_{\lambda}\bar{S}^{\lambda}$. These terms will be eliminated below by using the remaining field equations.
 
 Taking the global shift invariance of the action into account, the scalar field equation can be rewritten in terms of a Noether current $J^{\mu}$, namely,
\begin{equation}
\nabla^{(g)}_{\mu}J^{\mu}=0,
\end{equation}  
where 
\begin{equation}
J^{\mu}=\alpha v^{\mu}+\frac{\gamma}{\kappa^2}R^{(\mu\nu)}(\Gamma)v_{\nu}+\beta \bar{S}^{\mu}.
\label{pk}
\end{equation}
Using Eq.(\ref{eqrw1}), the current can be presented as:
\begin{eqnarray}
\label{current}
\nonumber J^{\mu}&=&\bigg\{\bigg[\alpha-\frac{\gamma}{(1+\gamma X)(2+3\gamma X)}(\gamma v^{\vartheta}v^{\lambda}T_{\vartheta\lambda}+T(1+\gamma X)+2(1+\gamma X)\beta v_{\lambda}\bar{S}^{\lambda})\bigg]g^{\mu\nu}+\\&+&\frac{\gamma}{(1+\gamma X)}T^{\mu\nu}\bigg\}v_{\nu}-\frac{\gamma}{(1+\gamma X)}v_{\lambda}v_{\vartheta}\epsilon^{\lambda}_{\,\,\beta\gamma\nu}S^{\vartheta\gamma\beta}g^{\mu\nu}+\beta \bar{S}^{\mu}.
\end{eqnarray}
This current will be used below.

\subsubsection{Connection equation}

Let us turn our attention to the connection equation. As a first step, for the sake of convenience, let us split it into its symmetric and antisymmetric pieces, respectively,
\begin{eqnarray}
\label{sym}\frac{1}{\sqrt{-h}}\nabla^{(C)}_{\alpha}[\sqrt{-h}h^{\mu\nu}]&=&S_{\alpha}h^{\mu\nu}-\frac{2}{3}S_{\rho}\delta^{(\mu}_{\alpha}h^{\nu)\rho};\\
\label{skew}0&=&-\frac{2}{3}S_{\rho}\delta^{[\nu}_{\alpha}h^{\mu]\rho}+2S^{[\nu}_{\ \alpha\lambda}h^{\mu]\lambda}+2\beta\kappa^{2}v_{\lambda}\frac{\epsilon^{\lambda\vartheta\nu\mu}}{\sqrt{1+\gamma X}}g_{\alpha\vartheta},
\end{eqnarray}
where we have used the connection decomposition, $\Gamma^{\mu}_{\nu\alpha}=C^{\mu}_{\nu\alpha}+S^{\mu}_{\nu\alpha}$, with $C^{\mu}_{\nu\alpha}=\Gamma^{\mu}_{(\nu\alpha)}$. The antisymmetric equation is non-dynamical, so it provides an algebraic relation between the torsion and the gradient of the scalar field. 
	
	The gauge freedom stemming from the projective invariance of the connection allows us to simplify further Eqs. (\ref{sym}, \ref{skew}) by means of a field redefinition. In fact, this is attainable by fixing a particular gauge, for instance, picking $A_{\mu}=-\frac{2}{3}S_{\mu}$ in Eq.(\ref{proj}), which leads to
	\begin{equation}
\label{gauge}	\tilde{\Gamma}^{\alpha}_{\mu\nu}=\Gamma^{\alpha}_{\mu\nu}-\frac{2}{3}\delta^{\alpha}_{\mu}S_{\nu}
	\end{equation}
	and, as a result,
	\begin{eqnarray}
\tilde{C}^{\alpha}_{\mu\nu}&=&C^{\alpha}_{\mu\nu}-\frac{1}{3}\left(\delta^{\alpha}_{\mu}S_{\nu}+\delta^{\alpha}_{\nu}S_{\mu}\right);\\
\tilde{S}^{\alpha}_{\mu\nu}&=&S^{\alpha}_{\mu\nu}-\frac{1}{3}\left(\delta^{\alpha}_{\mu}S_{\nu}-\delta^{\alpha}_{\nu}S_{\mu}\right).
\end{eqnarray}
This suggestive gauge choice allows us to gauge away the vector part of the torsion tensor; we mean, $\tilde{S}_{\alpha}$ vanishes. It is worth stressing that other gauge fixings could be chosen; however, the resulting effective on-shell action remains unchanged, as expected. So, the symmetric and antisymmetric pieces, in this gauge fixing, take a simpler form as displayed below:
\begin{align}
&\label{eqna}\nabla^{(\tilde{C})}_{\alpha}[\sqrt{-h}h^{\mu\nu}]=0;\\
&\label{TOC}2\tilde{S}^{[\nu}_{\ \alpha\lambda}h^{\mu]\lambda}+2\beta\kappa^{2}v_{\lambda}\frac{\epsilon^{\lambda\vartheta\nu\mu}}{\sqrt{1+\gamma X}}g_{\alpha\vartheta}=0.
\end{align}
One finds straightforwardly from Eq. (\ref{eqna}) that
\begin{equation}
\tilde{C}^{\alpha}_{\mu\nu}=\,^{(h)}L^{\alpha}_{\mu\nu},
\end{equation}
where 
\begin{equation}
\,^{(h)}L^{\alpha}_{\mu\nu}=\frac{1}{2}h^{\alpha\vartheta}\left(-\partial_{\vartheta}h_{\mu\nu}+\partial_{\mu}h_{\nu\vartheta}+\partial_{\nu}h_{\mu\vartheta}\right)
\end{equation}
 are simply the Christoffel symbols for the metric $h_{\mu\nu}$. As stated above, the anti-symmetric equation actually yields a constraint. To clarify this point, we contract the Eq. (\ref{TOC}) with $g_{\mu\eta}$ to arrive at
 \begin{equation}
 S_{[\eta\nu]\alpha}=\beta \kappa^{2}v_{\lambda}\epsilon^{\lambda}_{\,\,\alpha\eta\nu}+\gamma P_{\eta\nu\alpha},
 \label{eqsw}
 \end{equation}
 where $P_{\eta\nu\alpha}=v^{\lambda}v_{[\nu}S_{\eta]\alpha\lambda}$. Contracting Eq.(\ref{eqsw}) with $\epsilon^{\nu\alpha\eta\vartheta}$, one obtains
 \begin{equation}
 \bar{S}^{\vartheta}=-6\beta\kappa^2 v^{\vartheta}+P^{\vartheta},
 \end{equation}
where $P^{\vartheta}=\gamma \epsilon^{\nu\alpha\eta\vartheta}v^{\lambda}v_{\eta}\tilde{S}_{\nu\alpha\lambda}$. Interestingly, this equation tells us that the axial vector is split into two parts such that the first is parallel to $v^{\mu}$ and the second is perpendicular to $v_{\mu}$. It is straightforward to check that $P^{\vartheta}$ is a normal vector to $v^{\vartheta}$, that is, $P^{\mu}v_{\mu}=0$. Consequently, we have the important relation
 \begin{equation}
v_{\mu}\bar{S}^{\mu}=-6\beta \kappa^2 X,
\end{equation}
  which ought to be used to eliminate the dependence of the axial torsion on the \textit{r.h.s} of Eq.(\ref{mest}).

It is useful to obtain the relation between the metrics $h_{\mu\nu}$ and $g_{\mu\nu}$. To do it, we follow the way presented in \cite{Delhom:2019wcm}, that is, using the matrix representation of Eq.(\ref{hg}), we can rewrite it as
	\begin{equation}
	\sqrt{-h}\hat{h}^{-1}=\sqrt{-g}\hat{g}^{-1}\left(\hat{I}+\gamma \hat{vv}\right).
	\end{equation}
	Now, taking the determinant of the former equation, one gets the following relation: $\sqrt{-h}=\sqrt{-g} \sqrt{\left(1+\gamma X\right)}$, where we have used the identity $\det(\hat{I}+\gamma\hat{vv})=1+\gamma X$ (see \cite{HornJohnson}). Substituting this into Eq.(\ref{hg}), we find
	\begin{equation}
	h^{\mu\nu}=\frac{g^{\mu\nu}+\gamma v^{\mu}v^{\nu}}{\sqrt{1+\gamma X}},
	\end{equation}
	and, as a consequence,
	\begin{equation}
	h_{\mu\nu}=\sqrt{1+\gamma X}\left(g_{\mu\nu}-\frac{\gamma}{1+\gamma X}v_{\mu}v_{\nu}\right).
    \label{ewq}
	\end{equation}
     Both metrics are disformally related. By defining the quantity $Y=h^{\mu\nu}(\partial_{\mu}\vartheta)(\partial_{\nu}\vartheta)$, it is possible to invert the former equations, i.e.,
	\begin{eqnarray}
	\nonumber g^{\mu\nu}&=&\sqrt{1+\gamma X}\,h^{\mu\nu}-\frac{\gamma}{1+\gamma X}\tilde{v}^{\mu}\tilde{v}^{\nu};\\
    g_{\mu\nu}&=&\frac{1}{\sqrt{1+\gamma X}}\left(h_{\mu\nu}+\frac{\gamma}{\sqrt{1+\gamma X}}v_{\mu}v_{\nu}\right),
    \label{kps}
	\end{eqnarray}
	where $X$ must be interpreted as a function of $Y\equiv h^{\mu\nu}v_{\mu}v_{\nu}$, which are related by $Y=X\sqrt{1+\gamma X}$ and $\tilde{v}^{\mu}\equiv h^{\mu\nu}v_{\nu}$.
	With this at hand, we can reverse Eq. \eqref{TOC}. To do that, let us define $T_{\rho\alpha\lambda}\equiv h_{\rho\nu}\tilde{S}^{\nu}_{\,\,\alpha\lambda}$. Now, multiplying Eq. \eqref{TOC} by $h_{\nu\rho}h_{\mu\sigma}$, one finds
    \begin{equation}
        T_{\rho\alpha\sigma}-T_{\sigma\alpha\rho}=-2\beta\kappa^{2}v_{\lambda}\frac{\epsilon^{\lambda\phantom{a}\nu\mu}_{\phantom{a}\alpha}}{\sqrt{1+\gamma X}}h_{\nu\rho}h_{\mu\sigma}.
    \end{equation}
    Defining $F_{\alpha\rho\sigma}=T_{\rho\alpha\sigma}-T_{\sigma\alpha\rho}$, so
    \begin{eqnarray}     \nonumber F_{\alpha\rho\sigma}&=&-2\beta\kappa^{2}v_{\lambda}\frac{\epsilon^{\lambda\phantom{a}\nu\mu}_{\phantom{a}\alpha}}{\sqrt{1+\gamma X}}h_{\nu\rho}h_{\mu\sigma}\\ &=&-2\beta\kappa^{2}v_{\lambda}\epsilon^{\lambda\phantom{a}}_{\phantom{a}\alpha\rho\sigma}\sqrt{1+\gamma X}
    \end{eqnarray}
    which satisfies 
    \begin{eqnarray}
        \nonumber T_{\sigma\rho\alpha}&=&\frac{1}{2}\left(F_{\alpha\rho\sigma}+F_{\rho\sigma\alpha}-F_{\sigma\alpha\rho}\right)\\
        \nonumber &=&\beta\kappa^2 v_{\lambda}\frac{\epsilon^{\lambda\phantom{a}\nu\mu}_{\phantom{a}\sigma}}{\sqrt{1+\gamma X}}h_{\nu\rho}h_{\mu\alpha}\\ &=&\beta\kappa^{2}v_{\lambda}\epsilon^{\lambda\phantom{a}}_{\phantom{a}\sigma\rho\alpha}\sqrt{1+\gamma X}
    \end{eqnarray}
  
    Finally, from Eq. (\ref{TOC}) and using $\tilde{S}^{\nu}_{\,\,\alpha\lambda}=T_{\rho\alpha\lambda}h^{\rho\nu}$, we arrive at
	\begin{equation}
\tilde{S}^{\alpha}_{\mu\nu}=-\beta\kappa^{2}v_{\lambda}\epsilon^{\alpha\lambda}_{\,\,\,\,\,\,\,\mu\nu},
\label{edc}
	\end{equation}
	or, in terms of the metric $h_{\mu\nu}$, the torsion reads
	\begin{equation}
	\tilde{S}^{\alpha}_{\mu\nu}=-\frac{\beta\kappa^{2}}{\sqrt{1+\gamma X}}v_{\lambda}h^{\alpha\vartheta}h^{\lambda\sigma}\epsilon^{(h)}_{\vartheta\sigma\mu\nu},
    \label{er}
	\end{equation}
	where we have defined $\epsilon^{(h)}_{\vartheta\sigma\mu\nu}=\sqrt{-h}\,\varepsilon_{\vartheta\sigma\mu\nu}=\sqrt{1+\gamma X}\,\epsilon_{\vartheta\sigma\mu\nu}$. Note that the torsion tensor is entirely sourced by the scalar field $\vartheta$ and its derivative. As a consequence of Eq. \eqref{edc}, the axial piece of the torsion tensor is given by
    \begin{equation}
\bar{S}^{\mu}=\epsilon^{\mu\nu\alpha\beta}\tilde{S}_{\nu\alpha\beta}=-6\beta\kappa^2 v^{\mu}.
    \end{equation}
         Since the projective modes can be gauged away, the full connection is given by
         \label{axial}
	\begin{equation}
\tilde{\Gamma}^{\alpha}_{\mu\nu}=\,^{(h)}L^{\alpha}_{\mu\nu}-\frac{\beta\kappa^{2}}{\sqrt{1+\gamma X}}v_{\lambda}h^{\alpha\vartheta}h^{\lambda\sigma}\epsilon^{(h)}_{\vartheta\sigma\mu\nu}.
\label{con1}
	\end{equation}
This is a remarkable result since the torsion tensor is entirely sourced by the derivative of the scalar field and its norm, $v_{\mu}$ and $X$.

\subsubsection{Metric field equation}

Using $\tilde{\Gamma}^{\mu}_{\alpha\beta}=\,^{(h)}L^{\mu}_{\alpha\beta}+\tilde{S}^{\mu}_{\alpha\beta}$, the Riemann tensor can be decomposed into 
\begin{equation}
    R^{\alpha}_{\phantom{a}\beta\mu\nu}(\tilde{\Gamma})=R^{\alpha}_{\phantom{a}\beta\mu\nu}(h)+\nabla_{\mu}^{(h)}\tilde{S}^{\alpha}_{\nu\beta}-\nabla_{\nu}^{(h)}\tilde{S}^{\alpha}_{\mu\beta}+\tilde{S}^{\lambda}_{\nu\beta}\tilde{S}^{\alpha}_{\mu\lambda}-\tilde{S}^{\lambda}_{\mu\beta}\tilde{S}^{\alpha}_{\nu\lambda}.
    \label{eg}
\end{equation}
With this definition at hand, one can rewrite the field equation \eqref{me} in the $h$-metric frame. Let us check this step-by-step. First, from the definition \eqref{eg}, the Ricci tensor can be cast into the form,
\begin{equation}
     R_{\beta\nu}(\tilde{\Gamma})=R_{\beta\nu}(h)+\nabla_{\lambda}^{(h)}\tilde{S}^{\lambda}_{\nu\beta}-\nabla_{\nu}^{(h)}\tilde{S}^{\lambda}_{\lambda\beta}+\tilde{S}^{\gamma}_{\nu\beta}\tilde{S}^{\lambda}_{\lambda\gamma}-\tilde{S}^{\lambda}_{\gamma\beta}\tilde{S}^{\gamma}_{\nu\lambda}.
     \label{es}
\end{equation}
Now, plugging Eq.\eqref{er} into Eq. \eqref{es},
it is easy to see that  
\begin{equation}
\begin{split}
     R_{(\beta\nu)}(\tilde{\Gamma})&=R_{\beta\nu}(h)-\tilde{S}^{\lambda}_{\gamma(\beta}\tilde{S}^{\gamma}_{\nu)\lambda}\\
     &=R_{\beta\nu}(h)+\frac{2\beta^2\kappa^4}{1+\gamma X}\left(Yh_{\mu\nu}-v_\mu v_\nu\right),
     \end{split}
\end{equation}
where $\nabla_{\lambda}^{(h)}\tilde{S}^{\lambda}_{(\nu\beta)}=0$ and $\tilde{S}^{\lambda}_{\lambda\gamma}=0$, as a result of the antisymmetric properties of the Levi-Civita tensor and the torsion tensor. Second, note that the combination in brackets of Eq.\eqref{me} can be expressed as
\begin{equation}
    \begin{split}
        R(\tilde\Gamma)+\gamma v^\alpha v^\beta R_{\alpha\beta}(\tilde\Gamma)&=R_{\alpha\beta}(\tilde\Gamma)h^{\alpha\beta}\sqrt{1+\gamma X}\\
        &=\sqrt{1+\gamma X}\,\tilde{R}(h)+\frac{6\beta^2\kappa^4 Y}{\sqrt{1+\gamma X}},
    \end{split}
\end{equation}
where $\tilde{R}(h)=h^{\mu\nu}R_{\mu\nu}(h)$. The third term on the l.h.s. of Eq.\eqref{me} reduces to
\begin{equation}
    2\gamma v^{\beta}v_{(\mu}R_{\nu)\beta}(\tilde{\Gamma})=\frac{2\gamma}{\sqrt{1+\gamma X}} v^{\beta}v_{(\mu}R_{\nu)\beta}(h).
\end{equation}
The last term on the left-hand side of Eq.\eqref{me} can be set into the form
\begin{equation}
\begin{split}
-2\beta\kappa^{2}v_{\lambda}\epsilon^{\lambda}_{\,\,\,\beta\gamma(\mu}S_{\nu)}^{\,\,\,\beta\gamma}=\frac{4\beta^2\kappa^4}{1+\gamma X}\left(Y h_{\mu\nu}-v_{\mu}v_{\nu}\right).
    \end{split}
\end{equation}
Putting all information together, the field equation in the $h$-metric frame becomes 
\begin{equation}
    \begin{split}
        G_{\mu\nu}(h)-\frac{\gamma v_{\mu} v_{\nu}\tilde{R}(h)}{2\sqrt{1+\gamma X}}+\frac{2\gamma\tilde{v}^{\beta}v_{(\mu}R_{\nu)\beta}(h)}{\sqrt{1+\gamma X}}+\frac{3\beta^2\kappa^4 Y h_{\mu\nu}}{1+\gamma X}-3\beta^2\kappa^4 v_{\mu}v_{\nu}\left(\frac{2+\gamma X}{1+\gamma X}\right)=\kappa^2 T_{\mu\nu},
    \end{split}
\end{equation}
    where $T_{\mu\nu}=T^{(h)}_{\mu\nu}+T^{(m)}_{\mu\nu}$ with
    \begin{equation}
        T^{(h)}_{\mu\nu}=-\frac{\alpha}{1+\gamma X}\left[v_{\mu}v_{\nu}\left(1+\frac{\gamma X}{2}\right)-\frac{1}{2}h_{\mu\nu}Y\right]
    \end{equation}
    is the $\vartheta$-dependent part of the energy-momentum tensor.

\subsubsection{Scalar field equation}
    
	By taking into account the gauge fixing Eq.(\ref{gauge}) and inserting the axial piece of the torsion \eqref{axial} and the connection (\ref{con1}) into Eq. (\ref{sc}), the scalar field equation (\ref{sc}) becomes
	\begin{align}
&\label{scf}\left(\alpha-6\beta^{2}\kappa^2\right)\square^{(g)}\vartheta+\frac{\gamma}{\kappa^2}\nabla^{(g)}_{\mu}\left[v_{\sigma}R^{(\mu\sigma)}(h)\right]=0.
	\end{align}

Let us clarify the meaning of the ghost-free condition in the present construction. After
integrating out the connection and the axial torsion, the effective coefficient multiplying
the scalar kinetic term is \(\alpha-6\beta^2\kappa^2\). Therefore, the region
\(\alpha-6\beta^2\kappa^2>0\) corresponds to the standard healthy sign of the reduced
scalar-torsion kinetic sector, whereas \(\alpha-6\beta^2\kappa^2<0\) would lead to a
ghostlike sign. The special branch $\alpha=6\beta^2\kappa^2$
used below is a degenerate branch of the reduced equations: the ordinary scalar kinetic
term is exactly canceled by the torsion-induced contribution. For Ricci-flat geometries in $h$-metric frame, the remaining curvature-dependent part of the above equation also vanishes. In this sense, the branch is ghost-free in the reduced
scalar-torsion sector relevant for the exact solutions considered here. A full perturbative
stability analysis of the physical metric \(g_{\mu\nu}\), however, lies beyond the scope of
the present work.


 As we have already discussed, we consider the special case corresponding to $V(\vartheta)=0$. In this case, the action becomes invariant under shift transformation, $\vartheta\rightarrow \vartheta+a$, and, as a consequence, it admits a conserved Noether current defined by
\begin{equation}
J^{\mu}=\frac{1}{\sqrt{-g}}\frac{\delta S}{\delta (\partial_{\mu}\vartheta)},
\end{equation} 
with its explicit form given by Eq. \eqref{pk},\eqref{current}. After integrating out the connection and the axial piece of the torsion tensor, one finds
\begin{equation}
J^{\mu}=(\alpha-6\beta^2 \kappa^2)\frac{\tilde{v}^{\mu}}{\sqrt{1+\gamma X}}+\frac{\gamma}{\kappa^2}\left[\tilde{v}^{\beta}\tilde{R}^{\mu}_{\phantom{a}\beta}(h)-\frac{\gamma}{(1+\gamma X)^{3/2}}\tilde{v}^{\mu}\tilde{v}^{\alpha}\tilde{v}^{\beta}R_{\alpha\beta}(h)\right],
\label{Noet}
\end{equation}
where $\tilde{R}^{\mu}_{\phantom{a}\beta}(h)=h^{\mu\alpha}R_{\alpha\beta}(h)$.
Taking these definitions into account, the scalar field equation reduces to the conservation of the Noether current
\begin{equation}
\nabla_{\mu}^{(g)}J^{\mu}= \partial_{\mu}\left(\sqrt{-g}J^{\mu}\right)=\nabla^{(h)}_{\mu}\tilde{\mathcal{J}}^{\mu}=0,
\label{sc2}
\end{equation}
where $\tilde{\mathcal{J}}^{\mu}\equiv \dfrac{J^{\mu}}{\sqrt{1+\gamma X}}$ is the conserved current in the $h$-metric frame.
Such a current plays an important role in global shifted scalar-tensor theories. In this context, one may recall a no-hair theorem
 for Galileon theories, where it has been argued \cite{Hui:2012qt} that for static and spherically symmetric spacetimes, the only non-trivial component of the Noether current $\tilde{\mathcal{J}}^{r}$ must vanish everywhere to ensure regularity of the square norm of the Noether current $\tilde{\mathcal{J}}^{2}=h^{\mu\nu}\tilde{\mathcal{J}}_{\mu}\tilde{\mathcal{J}}_{\nu}$  at the horizon. They also argued that such a setup implies that the scalar field must be constant everywhere. In our particular case, this no-hair theorem  is recovered, namely, from Eq.(\ref{Noet}) it is straightforward to check that $\vartheta=const$ implies $J^{\mu}=0$ or, equivalently, $\tilde{\mathcal{J}}^{\mu}=0$. Nevertheless, in much the same way as what happens in \cite{Babichev:2013cya}, we have an alternative condition to keep $\tilde{\mathcal{J}}^2$ regular at the horizon for Ricci-flat black holes, i.e., by taking 
\begin{equation}
\alpha=6\beta^{2}\kappa^2, 
\label{al}
\end{equation}
one has $\tilde{\mathcal{J}}^{\mu}=0$ throughout spacetime without restricting  $\vartheta(r)$ to a constant everywhere, while still satisfying the no-hair regularity condition. As pointed out in \cite{Babichev:2013cya}, this alternative condition is more convenient because there are no constraints on the scalar field.

In the next section we shall explore static symmetric black holes in more detail following the aforementioned line of reasoning.


 \section{Applications: Exact Ricci-flat black-hole solutions}
 \label{sec3}

	In this section, let us study the consistency of some black holes within the shifted symmetric modified theory. Before to begin with, it is worthwhile to emphasize an important assumption that we shall proceed with, namely, in addition to imposing Eq. (\ref{al}), let us also impose regularity of the norm of $v_{\mu}$ at the horizon, which is easily reached by the requirement that $Y=h^{\mu\nu}v_{\mu}v_{\nu}=const$, even though $v_{\mu}$ does not.   
	This is not an arbitrary choice; rather, it is technically convenient because the field equations will be rather simplified. Fixed-norm vectors have been explored in the context of spontaneous Lorentz symmetry-breaking models; see, for example, \cite{Kostelecky:1989jw,Jacobson:2000xp}. Such a constraint can be incorporated from the very beginning by adding a Lagrange multiplier of the form 
\begin{equation}
S_{L}=\int d^{4}x\sqrt{-g}\, \sigma\, (v^{\mu}v_{\mu}-X_0)=\int d^{4}x\frac{\sqrt{-h}}{(1+\gamma X)}\, \sigma\, (\tilde{v}^{\mu}v_{\mu}-Y_0),
\end{equation}	
in the action (\ref{total}), with $\sigma$ being the Lagrange multiplier. Although this procedure is standard, it is more practical to apply the fixed-norm constraint directly to the field equations once we have already found them; as a result, we have a reminiscence of the Einstein-aether theory \cite{Jacobson:2000xp}.  

We explore a simple class of exact solutions of the theory. The key observation is that, after solving the connection equation, the field equations take a particularly transparent form in the $h$-metric frame. We use the notation $z\equiv \sqrt{1+\gamma X}$ for brevity. Consequently, the two kinetic terms are related by the following relation:
\begin{equation}
    Y=zX ,
\end{equation}
or, equivalently, by the cubic equation
\begin{equation}
    z^3-z-\gamma Y=0 .
\end{equation}

In the \(h\)-metric frame, the metric field equation can be written as
\begin{align}
    G_{\mu\nu}(h)
    &-
    \frac{\gamma}{2z}v_\mu v_\nu R(h)
    +
    \frac{2\gamma}{z}v_{(\mu}R_{\nu)\rho}(h)\tilde{v}^\rho
    \nonumber\\
    &+
    \frac{Y}{z^2}
    \left(
        3\beta^2\kappa^4-\frac{\alpha\kappa^2}{2}
    \right)h_{\mu\nu}
    -
    \frac{1+z^2}{z^2}
    \left(
        3\beta^2\kappa^4-\frac{\alpha\kappa^2}{2}
    \right)v_\mu v_\nu
    =
    0 .
    \label{eq:h-frame-field-eq}
\end{align}
Therefore, in the ghost-free branch
\begin{equation}
    \alpha=6\beta^2\kappa^2 ,
    \label{eq:ghost-free-branch}
\end{equation}
the algebraic scalar-torsion sector cancels identically. In this case, Eq.~\eqref{eq:h-frame-field-eq} reduces to
\begin{equation}
    G_{\mu\nu}(h)
    -
    \frac{\gamma}{2z}v_\mu v_\nu R(h)
    +
    \frac{2\gamma}{z}v_{(\mu}R_{\nu)\rho}(h)\tilde{v}^\rho
    =
    0 .
    \label{eq:h-frame-ghost-free}
\end{equation}
It follows immediately that every Ricci-flat geometry in the $h$-metric frame,
\begin{equation}
    R_{\mu\nu}(h)=0 ,
\end{equation}
solves the metric field equations on the branch \eqref{eq:ghost-free-branch}. Furthermore, the Noether current associated with the shift symmetry (\ref{current}) becomes
\begin{equation}
    \tilde{\mathcal{J}}^\mu
    =
    \frac{\alpha-6\beta^2\kappa^2}{z^2}\tilde{v}^\mu
    +
    \frac{\gamma }{\kappa^2 z}
    \left[
        \tilde{R}^\mu{}_\nu(h)\tilde{v}^\nu
        -
        \frac{\gamma}{z^3}
        \tilde{v}^\mu R_{\alpha\beta}(h)\tilde{v}^\alpha \tilde{v}^\beta
    \right].
    \label{eq:h-frame-current}
\end{equation}
Thus, for Ricci-flat \(h_{\mu\nu}\), and imposing the regularity condition of the norm of the Noether current 
\(\alpha=6\beta^2\kappa^2\), one finds
\begin{equation}
    \tilde{\mathcal{J}}^\mu=0 ,
\end{equation}
and the scalar equation is automatically satisfied for a non-trivial scalar field.

\subsection{Schwarzschild metric in the $h$-metric frame}

As a particular example of a spherically symmetric spacetime, let us take the metric \(h_{\mu\nu}\) to be the Schwarzschild one,
\begin{equation}
    ds_h^2
    =
    -A(r)dt^2
    +
    \frac{dr^2}{A(r)}
    +
    r^2d\Omega^2 ,
    \qquad
    A(r)=1-\frac{2M}{r}.
    \label{eq:h-schwarzschild}
\end{equation}
Since $R_{\mu\nu}(h)=0$, this geometry solves the field equations, in the $h$-metric frame, for the vanishing of the Noether current, $\alpha=6\beta^2 \kappa^2$. We now present two scalar profiles.

\subsubsection{Static fixed-norm profile}

A simple static profile is given by
\begin{equation}
    v_\mu=\left(0,\sqrt{\frac{q}{A(r)}},0,0\right),
    \label{eq:static-profile}
\end{equation}
where $q$ is a positive constant.
Then, imposing the fixed-norm condition, one has
\begin{equation}
    Y=h^{\mu\nu}v_\mu v_\nu=q.
    \label{scalar}
\end{equation}

Plugging Eq. \eqref{eq:h-schwarzschild} into Eq. \eqref{kps}, we find the line element in the $g$-metric frame is given by
\begin{equation}
    \dd s_g^2=-\frac{A(r)}{z}\dd t^2+\frac{z}{A(r)}\dd r^2+\frac{r^2}{z}\dd\Omega^2.
    \label{eq:g-static-schwarzschild}
\end{equation}
The scalar field is obtained from Eq. \eqref{scalar} by substituting the vector  \eqref{eq:static-profile} into the condition \eqref{scalar}; thus, one finds
\begin{equation}
    \vartheta(r)=\vartheta_0\pm \sqrt{q}\int\frac{\dd  r}{\sqrt{1-2M/r}},
\end{equation}
which can be easily integrated to get
\begin{equation}    \vartheta(r)=\vartheta_0\pm\sqrt{q}\left[\sqrt{r(r-2M)}+2M\ln\left(\sqrt r+\sqrt{r-2M}\right)\right].
    \label{eq:static-scalar}
\end{equation}
It is noteworthy that the scalar field is finite at the horizon $r=2M$. Its radial coordinate derivative diverges in Schwarzschild coordinates, but the invariant norms $Y=q$ and $X=q/z$ are finite. 
Let us emphasize the interpretation of this solution. Since the $h_{\mu\nu}$ metric is
Schwarzschild, \(R_{\mu\nu}(h)=0\), and we are working on the branch
\(\alpha=6\beta^2\kappa^2\), the \(h\)-frame metric equations are satisfied even though the
scalar field is nontrivial. Moreover, the Noether current vanishes identically on this branch.
Therefore the scalar profile is stealth with respect to \(h_{\mu\nu}\) for a discussion of stealth solutions, see e.g. \cite{Ayon-Beato:2004nzi}:
the scalar is present, but the background remains a Schwarzschild spacetime in the $h$-metric frame.

This interpretation as a stealth solution, however, applies only in the \(h\)-frame. The metric $g_{\mu\nu}$
is disformally related to \(h_{\mu\nu}\), namely
\[
g_{\mu\nu}=\frac{1}{z}h_{\mu\nu}+\frac{\gamma}{z^2}\partial_\mu\vartheta\partial_\nu\vartheta ,
\]
and therefore the scalar gradient deforms the physical geometry whenever \(\gamma\neq0\).
Thus the solution should not be interpreted as a purely stealth black hole in the physical
frame. Instead, it represents a Schwarzschild background (in the $h$-metric frame) dressed by a secondary
disformal scalar hair in the physical \(g\)-frame.

The Kretschmann scalar invariant for the metric ~\eqref{eq:g-static-schwarzschild} is
\begin{equation}
    K_g\equiv R_{\alpha\beta\mu\nu}(g)R^{\alpha\beta\mu\nu}(g)
    =\frac{4(z^2-1)^2}{z^2r^4}+\frac{16M(z^2-1)}{z^2r^5}+\frac{48M^2}{z^2r^6}.
    \label{eq:kretschmann}
\end{equation}
It shows that the divergence at $r=2M$ of the radial coordinate derivative is only a problem of a bad choice of the coordinate system. Note also that for $z=1$ one recovers the Schwarzschild invariant $48M^2/r^6$.

The physical metric \eqref{eq:g-static-schwarzschild} can be viewed as a black hole with a deficit solid angle. It can be seen more clearly by proceeding with the following coordinate redefinitions, namely: $\bar{t}\to \dfrac{t}{\sqrt{z}}$, $\bar{r}\to r\sqrt{z}$. With this redefinition, the metric \eqref{eq:g-static-schwarzschild} becomes
\begin{equation}
    \dd s_g^2=-\left(1-\frac{2M_{eff}}{\bar{r}}\right)\dd \bar{t}^2+\frac{1}{1-\frac{2M_{eff}}{\bar{r}}}\dd \bar{r}^2+\frac{\bar{r}^2}{z^2}\dd\Omega^2,
\end{equation}
where we define the effective Schwarzschild mass $M_{eff}=M\sqrt{z}$.
Note that the area of the $2$-sphere is now $4\pi \left(\dfrac{r}{z}\right)^2$. The deficit solid angle is then defined by 
\begin{equation}
    \Delta\Omega=4\pi\left(1-\frac{1}{z^2}\right).
\end{equation}
Thus, the line element in the $g$-metric frame takes the form
\begin{equation}
     \dd s_g^2=-\left(1-\frac{2M_{eff}}{\bar{r}}\right)\dd \bar{t}^2+\frac{1}{1-\frac{2M_{eff}}{\bar{r}}}\dd \bar{r}^2+(1-\Delta)\bar{r}^2\dd\Omega^2,
\end{equation}
where $\Delta=\left(1-\dfrac{1}{z^2}\right)$.

Although the physical metric displays a solid angle deficit of the same geometrical type as the Barriola--Vilenkin global monopole spacetime \cite{Barriola:1989hx}, the origin of this deficit is different. In the present model there is no scalar triplet, and no topological charge. The deficit is instead generated by the disformal relation between the physical metric $g_{\mu\nu}$ and the Ricci-flat metric $h_{\mu\nu}$, through the scalar-gradient norm entering the algebraic factor $z$.

\subsubsection{Time-dependent regular scalar profile}
\label{sec:time-dependent}

We now allow for the scalar field to be time-dependent. We choose a similar ansatz for the scalar field profile proposed in \cite{Babichev:2013cya}, so
 \begin{equation}
    \vartheta(t, r)
    =    qt+\psi(r),
    \label{eq:schwarzschild-v-profile}
\end{equation}
where $q$ is a positive constant. This choice leads to 
\begin{equation}
v_{\mu}=[q,  \psi^{\prime}(r),0,0],
\end{equation}
where the prime is for derivative with respect to the radial coordinate.
Now, let us impose the fixed-norm condition, 
\begin{equation}
    Y=h^{\mu\nu}v_\mu v_\nu=-\frac{q^2}{A(r)}+A(r)(\psi^{\prime}(r))^2=Y_{0},
    \label{kk}
\end{equation}
where $Y_{0}$ is a constant.
To check the regularity condition of the scalar field at the horizon, one rewrites Eq. \eqref{kk} as
\begin{equation}
\psi^{\prime}(r)=\pm \frac{\sqrt{q^2+Y_0 A(r)}}{A(r)}.
\label{pse}
\end{equation}
Note that near the horizon, $A(r)\to0$, the equation for $\psi^{\prime}(r)$ behaves as
\begin{equation}
\psi^{\prime}(r)=\pm \frac{|q|}{A(r)}+...\,\,,
\end{equation} 
where the ellipses stand for subleading terms. At first sight, this seems to diverge at the horizon. However, it reflects the "bad" choice of coordinates. To check regularity at the future horizon, introduce the advanced Eddington–Finkelstein coordinate
\begin{equation}
v=t+r_{*}=t+r+2M\ln\left(\frac{r}{2M}-1\right),
\label{kjv}
\end{equation} 
where $r_{*}$ is the tortoise coordinate. Using the advanced coordinate,
\begin{equation}
    \vartheta(v,r)=q\left(v-r-2M\ln\left(\frac{r}{2M}-1\right)\right)+\psi(r).
\end{equation}
Therefore, for the positive branch, near the horizon, we have
\begin{equation}
    \vartheta_+ (v,r)=qv+ \mbox{finite radial function}.
\end{equation}
This shows that the scalar field is regular at the future horizon.

After using the relation \eqref{kps}, the line element in the $g$-metric frame reads
\begin{equation}
    \dd s_g^2=\frac1z\dd s_h^2+\frac{\gamma}{z^2}\left[q\dd t+\psi'(r)\dd r\right]^2.
    \label{eq:g-time-general}
\end{equation}
Thus, explicitly,
\begin{align}
    g_{tt}&=-\frac{A(r)}{z}+\frac{\gamma q^2}{z^2},\\
    g_{tr}&=\frac{\gamma q\psi'}{z^2},\\
    g_{rr}&=\frac1{zA(r)}+\frac{\gamma\psi'^2}{z^2},\\
    g_{\vartheta\vartheta}&=\frac{r^2}{z},\qquad
    g_{\phi\phi}=\frac{r^2}{z}\sin^2\vartheta.
\end{align}
The off-diagonal component $g_{tr}$ is a radial shift induced by the time-dependent scalar; it is not a rotating term.

A particularly close analogue of the stealth Schwarzschild black hole found in \cite{Babichev:2013cya} is obtained by choosing
\begin{equation}
    Y_0=-q^2.
\end{equation}
Substituting it into Eq.\eqref{pse}, we obtain
\begin{equation}
    \psi'(r)=\pm\frac{q\sqrt{1-A(r)}}{A(r)}
    =\pm\frac{q\sqrt{2M/r}}{1-2M/r}.
\end{equation}
Now, integrating it and using Eq.\eqref{eq:schwarzschild-v-profile}, we arrive at 
\begin{equation}
    \vartheta_\pm(t,r)=qt\pm q\left[2\sqrt{2Mr}+2M\ln\left(\frac{\sqrt r-\sqrt{2M}}{\sqrt r+\sqrt{2M}}\right)\right]+\vartheta_0,
    \label{eq:theta-bc-like}
\end{equation}
where $\vartheta_0$ and $q$ are constants.
Both solutions above (with plus and minus sign) coincide with that found by Babichev and Charmousis \cite{Babichev:2013cya} with the identification $\mu=2M$. To further verify the regularity at the horizon, let us use the advanced coordinate \eqref{kjv}; in this case, the positive branch of the solution is
\begin{equation}
    \vartheta_+(v,r)=q\left[v-r+2\sqrt{2Mr}-4M\ln\left(1+\sqrt{\frac{r}{2M}}\right)\right]+\vartheta_0,
    \label{eq:theta-ef}
\end{equation}
clearly confirming its regularity at $r=2M$. Furthermore, the gradient of the scalar field, $v_{\mu}$, is also regular at the horizon. The line element in the $g$-metric frame, in regular coordinates, reads
\begin{equation}
    \dd s_g^2=\frac1z\left[-A(r)\dd v^2+2\dd v\dd r+r^2\dd\Omega^2\right]
    +\frac{\gamma}{z^2}\left[q\dd v+\left(\psi'(r)-\frac{q}{A(r)}\right)\dd r\right]^2.
    \label{eq:g-ef}
\end{equation}
For the future-regular branch,
\begin{equation}
    \psi'(r)-\frac{q}{A(r)}=\frac{\sqrt{q^2+Y_0 A(r)}-q}{A(r)}\to \frac{Y_0}{2q},\quad \mbox{when}\quad r\to 2M,
\end{equation}
so Eq.~\eqref{eq:g-ef} is regular at the future horizon.

Therefore, from the perspective of the $h$-frame, we constructed a stealth black hole with a non-trivial scalar field (regular everywhere), with $\mathcal{J}^{\mu}=0$ and $\mathcal{J}^2=0$. On the other hand, from the perspective of the $g$-metric frame, we constructed a black hole setup with secondary hair because matter is assumed to couple to the physical metric $g_{\mu\nu}$, and $g_{\mu\nu}$ is disformally related to $h_{\mu\nu}$; the scalar gradient induces a deformation of the physical geometry. Hence the solution (as we have seen before) is better interpreted as carrying secondary disformal scalar hair rather than as a stealth configuration in the physical frame.

\subsection{Kerr metric in the $h$-metric frame}

We now consider the case in which the metric is a Kerr background in the $h$-metric frame. We also denote the Kerr rotation parameter by \(a_K\). In
Boyer--Lindquist coordinates, the line element is
\begin{align}
ds_h^2 =&
-\left(1-\frac{2Mr}{\Sigma}\right)dt^2
-\frac{4Ma_K r\sin^2\theta}{\Sigma}\,dt\,d\phi
+\frac{\Sigma}{\Delta}\,dr^2+\Sigma\,d\theta^2
+\frac{\mathcal{A}\sin^2\theta}{\Sigma}\,d\phi^2 ,
\end{align}
where
\begin{align}
\Sigma=r^2+a_K^2\cos^2\theta, \qquad
\Delta=r^2-2Mr+a_K^2 ,
\end{align}
and
\begin{align}
\mathcal{A}=(r^2+a_K^2)^2-a_K^2\Delta\sin^2\theta .
\end{align}
Since the Kerr metric is Ricci-flat,
\begin{align}
R_{\mu\nu}(h)=0,
\end{align}
it solves the \(h\)-frame metric equations on the ghost-free branch
\begin{align}
\alpha=6\beta^2\kappa^2 .
\end{align}
On the same branch, the Noether current vanishes identically for Ricci-flat
\(h_{\mu\nu}\), and therefore the scalar field equation is automatically satisfied.
Thus Kerr provides another exact black-hole geometry of the theory in the $h$-metric frame.
It should be emphasized that the Kerr character of the solution refers to the
metric \(h_{\mu\nu}\). The metric \(g_{\mu\nu}\) is obtained through the disformal
map \eqref{kps}.
Consequently, for \(\gamma\neq0\) and a nontrivial scalar gradient, the physical geometry is
not the Kerr metric, but a Kerr-derived disformal geometry. Therefore, the usual Kerr
horizon, ergosurface, separability properties, and Carter constant cannot be assumed a
priori for \(g_{\mu\nu}\). They must instead be determined directly from the physical metric.
In the present work, we establish the existence of the exact Kerr solution in the $h$-metric frame and its
associated disformal physical geometry; a detailed analysis of the causal structure and
geodesic separability of \(g_{\mu\nu}\) will be carried out elsewhere.

Regular time-dependent stealth Kerr configurations with constant scalar kinetic term have been studied in higher-order scalar-tensor theories in \cite{Charmousis:2019vnf,Takahashi:2020hso}. In particular, the asymptotically flat limit of the constant-kinetic-term construction constrains the globally regular scalar profile to have no angular dependence. Disformal transformations of stealth Kerr solutions and the resulting non-Kerr rotating geometries have also been investigated in ~\cite{BenAchour:2020fgy,Anson:2020trg}, including their causal and geodesic properties and the breaking of the usual Kerr hidden symmetries. These results provide a useful point of comparison for the physical $g_{\mu\nu}$ geometry obtained below.

\subsubsection{Static fixed-norm profile and axis regularity}

Let us first construct the direct analogue of the static Schwarzschild profile. We impose
the constant \(h\)-norm condition
\begin{align}
Y \equiv h^{\mu\nu}\partial_\mu\vartheta\,\partial_\nu\vartheta = Y_0 ,
\end{align}
where \(Y_0\) is constant, and take the separable ansatz
\begin{align}
\vartheta(r,\theta)=R(r)+\Theta(\theta).
\end{align}
Using the inverse Kerr metric, the fixed-norm condition becomes
\begin{align}
\Delta \left(R'(r)\right)^2+\left(\Theta'(\theta)\right)^2
=
Y_0\left(r^2+a_K^2\cos^2\theta\right).
\end{align}
Introducing a separation constant \(C\), one obtains
\begin{align}
\Delta \left(R'(r)\right)^2 &= Y_0 r^2+C,\\
\left(\Theta'(\theta)\right)^2 &= Y_0 a_K^2\cos^2\theta-C.
\label{LK}
\end{align}

In order to have regularity and a smooth axisymmetric scalar field, one requires 
\begin{align}
\Theta'(0)=\Theta'(\pi)=0.
\label{constraint}
\end{align}
Applying this constraint to the angular equation \eqref{LK}, we get
\begin{align}
C=Y_0a_K^2.
\end{align}
Putting this result back into Eq.\eqref{LK}, one finds
\begin{align}
\left(\Theta'(\theta)\right)^2
=-Y_0a_K^2\sin^2\theta.
\end{align}
For a rotating configuration, $a_K\neq0$, reality of the scalar profile on the full angular interval requires
\begin{align}
Y_0\leq0.
\end{align}
On the other hand, using the constraint \eqref{constraint},  the radial equation becomes
\begin{align}
\Delta\left(R'(r)\right)^2=Y_0\left(r^2+a_K^2\right).
\end{align}
Since $\Delta>0$ in the Kerr exterior and, in particular, the large-$r$ behavior must remain real, the radial sector requires
\begin{align}
Y_0\geq0.
\end{align}
Consequently,
\begin{align}
Y_0=0,\qquad C=0,
\end{align}
and hence
\begin{align}
R'(r)=0,\qquad \Theta'(\theta)=0.
\end{align}
Thus the static separated ansatz does not yield a nontrivial scalar configuration that is simultaneously real throughout the exterior and smooth on the full rotation axis. The only globally regular static solution within this ansatz is a constant scalar field. In this case $v_\mu=0$, $z=1$, and the two metrics coincide, $g_{\mu\nu}=h_{\mu\nu}$, so that no nontrivial static disformal Kerr hair is generated.

\subsubsection{Stationary future-regular profile}

As in the Schwarzschild case, one may construct a scalar profile with linear time
dependence while keeping the background stationary in the $h$-metric frame. This is the Kerr analogue of
the Babichev--Charmousis-type scalar configuration. We take
\begin{align}
\vartheta(t,r,\theta)=qt+R(r)+\Theta(\theta),
\end{align}
where \(q\) is a constant. Since the theory is shift-symmetric, the physical metric depends
only on the one-form \(d\vartheta\). Therefore, although the scalar itself depends linearly
on \(t\), the disformally related metric is stationary.

The fixed-norm condition,
\begin{align}
Y=h^{\mu\nu}\partial_\mu\vartheta\,\partial_\nu\vartheta=Y_0,
\end{align}
gives, after multiplying by \(\Sigma\),
\begin{align}
Y_0\Sigma
=
\Delta \left(R'\right)^2
+
\left(\Theta'\right)^2
-
\frac{q^2(r^2+a_K^2)^2}{\Delta}
+
a_K^2q^2\sin^2\theta .
\end{align}
This equation can be separated. Introducing a separation constant \(K\), we write
\begin{align}
\Delta \left(R'\right)^2
-
\frac{q^2(r^2+a_K^2)^2}{\Delta}
-
Y_0r^2
&=
-K,
\\
\left(\Theta'\right)^2
+
a_K^2q^2\sin^2\theta
-
Y_0a_K^2\cos^2\theta
&=
K .
\end{align}
Hence,
\begin{align}
R'_{\pm}(r)
=
\pm
\frac{
\sqrt{
q^2(r^2+a_K^2)^2+\Delta\left(Y_0r^2-K\right)
}
}{\Delta},
\label{radialeq}
\end{align}
and
\begin{align}
\Theta'_{\pm}(\theta)
=
\pm
\sqrt{
K+Y_0a_K^2\cos^2\theta-a_K^2q^2\sin^2\theta
}.
\label{angulareq}
\end{align}

As discussed before, global regularity on the rotation axis imposes the constraint \eqref{constraint}, 
which, by applying it to the angular equation \eqref{angulareq}, provides
\begin{align}
K=-Y_0a_K^2.
\end{align}
Thus, the angular equation \eqref{angulareq} becomes
\begin{align}
\left(\Theta'(\theta)\right)^2
=-a_K^2\left(Y_0+q^2\right)\sin^2\theta.
\end{align}
Reality on the full angular interval requires
\begin{align}
Y_0+q^2\leq0.
\end{align}
At the same time, the radicand in the radial equation \eqref{radialeq} behaves for large $r$ as
\begin{align}
q^2(r^2+a_K^2)^2+\Delta\left(Y_0r^2-K\right)
=\left(q^2+Y_0\right)r^4+\mathcal{O}(r^3),
\end{align}
so reality of the radial profile at asymptotic infinity requires
\begin{align}
Y_0+q^2\geq0.
\end{align}
The two conditions are compatible if and only if
\begin{align}
Y_0=-q^2,\qquad K=a_K^2q^2,
\end{align}
which implies
\begin{align}
\Theta'(\theta)=0
\end{align}
throughout the spacetime. Thus the globally regular asymptotically flat stationary solution is nontrivial but has no angular dependence. This is consistent with the asymptotically flat constant-kinetic-term stealth Kerr profile discussed in ~\cite{Charmousis:2019vnf,Takahashi:2020hso}.

Given all that, the radial equation \eqref{radialeq} drastically simplifies to
\begin{align}
R'_{\pm}(r)
=
\pm |q|\,
\frac{\sqrt{2Mr\left(r^2+a_K^2\right)}}{\Delta}.
\end{align}
Let us now analyze regularity at the future event horizon. For a non-extremal Kerr black hole,
\begin{align}
r_+=M+\sqrt{M^2-a_K^2},
\end{align}
and we introduce ingoing Kerr coordinates,
\begin{align}
dv=dt+\frac{r^2+a_K^2}{\Delta}\,dr,
\qquad
d\tilde{\phi}=d\phi+\frac{a_K}{\Delta}\,dr .
\end{align}
The scalar one-form becomes
\begin{align}
d\vartheta
=
q\,dv
+
\left[
R'(r)-q\frac{r^2+a_K^2}{\Delta}
\right]dr.
\end{align}
The branch regular on the future horizon is obtained by taking
\begin{align}
R'_{\rm reg}(r)
=
q\,\frac{\sqrt{2Mr\left(r^2+a_K^2\right)}}{\Delta}.
\end{align}
Indeed,
\begin{align}
R'_{\rm reg}(r)-q\frac{r^2+a_K^2}{\Delta}
=
q\,\frac{\sqrt{2Mr\left(r^2+a_K^2\right)}-\left(r^2+a_K^2\right)}{\Delta},
\end{align}
and the numerator vanishes at $r=r_+$ because
\begin{align}
2Mr_+=r_+^2+a_K^2.
\end{align}
More explicitly,
\begin{align}
R'_{\rm reg}(r)-q\frac{r^2+a_K^2}{\Delta}
\longrightarrow -\frac{q}{2},
\qquad r\to r_+,
\end{align}
so the scalar one-form is finite at the future horizon. The regular scalar profile can be written as
\begin{align}
\vartheta_{\rm reg}(v,r)
=
qv
+
q\int dr\,
\frac{\sqrt{2Mr\left(r^2+a_K^2\right)}-\left(r^2+a_K^2\right)}{\Delta(r)}
+\vartheta_0.
\end{align}

In ingoing Kerr coordinates, the Kerr line element in the $h$-metric frame takes the regular form
\begin{align}
ds^2_{h}
=&
-\left(1-\frac{2Mr}{\Sigma}\right)dv^2
+2\,dv\,dr
+\Sigma\,d\theta^2
-\frac{4Ma_Kr\sin^2\theta}{\Sigma}\,dv\,d\tilde{\phi}
\nonumber\\
&\quad
-2a_K\sin^2\theta\,dr\,d\tilde{\phi}
+
\frac{\mathcal{A}\sin^2\theta}{\Sigma}\,d\tilde{\phi}^{\,2}.
\end{align}
The $g$-metric frame, the line element is therefore
\begin{align}
ds^2_g
=
\frac{1}{z}ds^2_{h}
+
\frac{\gamma q^2}{z^2}
\left\{
dv
+
\frac{\sqrt{2Mr\left(r^2+a_K^2\right)}-\left(r^2+a_K^2\right)}{\Delta}
\,dr
\right\}^2,
\end{align}
which is manifestly regular at the future horizon. Since the scalar has no angular dependence, there is no scalar-induced $r$--$\theta$ component in $g$-metric frame.

The resulting $g$-metric is closely related to the disformal Kerr geometries studied in ~\cite{BenAchour:2020fgy,Anson:2020trg}, where disformal transformations of regular constant-kinetic-term stealth Kerr configurations were used to generate rotating non-Kerr solutions. Those works also analyzed characteristic geometrical and geodesic properties of the disformed spacetime, including non-circularity and the breaking of the usual Kerr hidden symmetries. In the present construction, however, the disformal relation arises after solving the independent connection of the projectively invariant metric-affine scalar-torsion theory, with the Ricci-flat Kerr geometry appearing naturally in the auxiliary $h_{\mu\nu}$ frame.

It is also useful to check the Schwarzschild limit. Taking $a_K\to0$ gives $K\to0$ and
\begin{align}
R'_{\rm reg}(r)
\longrightarrow
q\,\frac{\sqrt{2M/r}}{A(r)},
\qquad
A(r)=1-\frac{2M}{r},
\end{align}
which is precisely the future-regular Schwarzschild branch of the previous subsection for the special value $Y_0=-q^2$ selected by global regularity in the rotating case.

Therefore, the stationary Kerr solution is stealth only in the $h$-metric frame. On the branch $\alpha=6\beta^2\kappa^2$, Kerr is Ricci-flat, the Noether current vanishes, and the scalar equation is automatically satisfied. In the  $g$-metric frame, however, the nontrivial scalar gradient produces a disformal deformation. The resulting configuration is thus naturally interpreted as a Kerr black hole with secondary disformal scalar hair in the $g$-metric frame, rather than as a purely stealth Kerr black hole in that frame.

\section{Conclusions}
\label{conc}

In this work, we constructed a projectively invariant ghost-free metric-affine scalar-torsion model in
which a pseudoscalar field is coupled to the Nieh--Yan density. After fixing the projective gauge, the independent connection
can be solved exactly. Its symmetric part becomes the Levi-Civita connection associated
with the \(h\)-metric frame, while its antisymmetric part is a purely axial torsion sourced by
the scalar gradient. The \(g\)-metric frame is then related to the \(h\)-metric frame
through a disformal transformation.

We showed that, after integrating out the connection and the axial torsion, the scalar
sector admits a special degenerate branch in which the ordinary scalar kinetic contribution
is canceled by the torsion-induced term. On this branch, Ricci-flat geometries in the
\(h\)-metric frame solves the metric field equations, and the Noether current vanishes
identically. Therefore, the scalar field equation is automatically satisfied without forcing
the scalar profile to be trivial. This provides a simple mechanism for constructing exact
black-hole solutions with nontrivial scalar configurations.

We applied this construction to Schwarzschild and Kerr geometries in the \(h\)-metric
frame. In the Schwarzschild case, we obtained both a static fixed-norm scalar profile and a
Babichev-Charmousis-like time-dependent profile. The latter is regular on the future
horizon when written in advanced Eddington-Finkelstein coordinates. The corresponding
geometry in the \(g\)-metric frame is not  Schwarzschild when the disformal coupling
is nonzero. In the static case, it can be interpreted as a Schwarzschild-like geometry with a
solid-angle deficit generated by the scalar-gradient sector, rather than by a
topological monopole charge.

In the Kerr case, regularity on the full rotation axis provides an essential additional constraint. For the static separated fixed-norm ansatz, smoothness at $\theta=0,\pi$ together with reality of both the angular and radial sectors forces $Y_0=C=0$, so the only globally regular static profile is trivial. For the stationary time-dependent configuration, the same axis regularity condition, combined with reality at large radius, instead selects a non-trivial combination of the parameters, $Y_0$ and $K$.
Thus a nontrivial rotating scalar configuration survives, but its globally regular asymptotically flat profile has no angular dependence. We constructed this branch explicitly in ingoing Kerr coordinates and showed that both the scalar one-form and the associated $g$-metric are regular at the future horizon. This restriction is consistent with previous constant-kinetic-term stealth Kerr constructions~\cite{Charmousis:2019vnf,Takahashi:2020hso}.

The black hole solutions accompanied by nontrivial scalar field profiles are stealth only in the $h$-metric frame. In the $g$-metric frame, matter couples to the disformally related metric, and the scalar gradient generically deforms the geometry. The resulting black holes should therefore be interpreted as carrying secondary disformal scalar hair rather than primary Noether-charge hair. In the Kerr case, the geometry in the $h$-metric frame is exactly Kerr, whereas the physical $g$-metric is Kerr-derived and disformally deformed. Closely related disformal Kerr geometries have been studied in ~\cite{BenAchour:2020fgy,Anson:2020trg}, including their causal structure and geodesic properties. A dedicated analysis of these properties within the present metric-affine realization is left for future work.

\acknowledgments
 E. Brito would like to thank UFPB, Universidade Federal da Paraíba, for its hospitality while part of this work
was carried out. The work of A. Yu. P. was partially supported by the Conselho Nacional de Desenvolvimento Cient\'ifico e Tecnol\'ogico (CNPq), project No.~303777/2023-0. 
P. J. Porf\'irio has been partially supported by the CNPq project No. 
307011/2026-7.


\end{document}